\documentclass[preprint, journal, dvipsnames, table]{vgtc}

\ifpdf
  \pdfoutput=1\relax                   
  \usepackage{graphicx}                
  \DeclareGraphicsExtensions{.pdf,.png,.jpg,.jpeg} 
\else
  \usepackage{graphicx}                
  \DeclareGraphicsExtensions{.eps}     
\fi%

\graphicspath{{figures/}{pictures/}{images/}{./}} 

\usepackage{microtype}                 
\PassOptionsToPackage{warn}{textcomp}  
\usepackage{textcomp}                  
\usepackage{mathptmx}                  
\usepackage{times}                     
\usepackage{cite}                      
\usepackage{tabu}                      
\usepackage{booktabs}                  

\usepackage[most]{tcolorbox}
\usepackage{pifont}
\usepackage{wrapfig}
\usepackage{tabularx}
\usepackage{hhline}
\usepackage{fontawesome}

\usepackage[frozencache]{minted}
\usepackage[inline,shortlabels]{enumitem}
\usepackage[skip=6pt]{caption}
\usepackage[normalem]{ulem}

\usepackage[english]{babel}
\usepackage{hyperref}   
\addto\extrasenglish{

}

\usepackage{booktabs}

\usepackage{makecell}
\usepackage{multirow}

\usepackage[most]{tcolorbox}
\usepackage{pifont}
\usepackage{tabularx}

\usepackage{calc}
\newlength\myheight
\newlength\mydepth
\settototalheight\myheight{Xygp}
\definecolor{user}{HTML}{d9ead3}
\definecolor{knowledge}{HTML}{fff2cc}
\definecolor{system}{HTML}{c9daf8}
\definecolor{systemDark}{HTML}{5c739c}
\definecolor{designer}{HTML}{d9d2e9}

\usepackage{tikz}

\newcommand{\component}[3]{
    \begin{tcolorbox}[skin=bicolor,fonttitle=\bfseries,coltitle=black,colbacktitle=#2,colback=#2!20,colframe=#2,title=#1, after skip=0.35em,left=3pt, right=3pt, top=3pt, bottom=3pt,boxsep=0pt,colbacklower=#2!10,middle=0.4em,toptitle=2pt, bottomtitle=2pt,sharp corners=all, boxrule=1pt, leftrule=0mm,breakable]
    #3
    \end{tcolorbox}%
}

\usepackage{xspace}

\newcommand{\yaml}[1]{\mintinline{yaml}{#1}}
\newcommand{\python}[1]{\mintinline{python}{#1}}
\newcommand{\lotse}{\mintinline{yaml}{Lotse}}

\preprinttext{To appear in IEEE Transactions on Visualization and Computer Graphics.}
\onlineid{0}

\vgtccategory{Research}
\vgtcpapertype{system}

\title{Lotse: A Practical Framework for Guidance in Visual Analytics}

\author{Fabian Sperrle, Davide Ceneda, and Mennatallah El-Assady}
\authorfooter{
\item Fabian Sperrle is with the University of Konstanz. Email: fabian.sperrle@uni-konstanz.de
\item Davide Ceneda is with TU Wien. Email: davide.ceneda@tuwien.ac.at
\item Mennatallah El-Assady is with the ETH AI Center. Email: menna.elassady@ai.ethz.ch
}

\shortauthortitle{Sperrle \MakeLowercase{\textit{et al.}}: Lotse: A Practical Framework for Guidance in Visual Analytics}

\abstract{Co-adaptive guidance aims to enable efficient human-machine collaboration in visual analytics, as proposed by multiple theoretical frameworks. This paper bridges the gap between such conceptual frameworks and practical implementation by introducing an accessible model of guidance and an accompanying guidance library, mapping theory into practice. We contribute a model of system-provided guidance based on  design templates and derived  strategies. We instantiate the model in a library called \lotse{} that allows specifying guidance strategies in definition files and generates running code from them. \lotse{} is the first guidance library using such an approach. It supports the creation of reusable guidance strategies to retrofit existing applications with guidance and fosters the creation of general guidance strategy patterns. We demonstrate its effectiveness through first-use case studies with VA researchers of varying guidance design expertise and find that they are able to effectively and quickly implement guidance with \lotse{}. Further, we analyze our framework's cognitive dimensions to evaluate its expressiveness and outline a summary of open research questions for aligning guidance practice with its intricate theory.} 

\teaser{
  \centering
  \includegraphics[width=\linewidth]{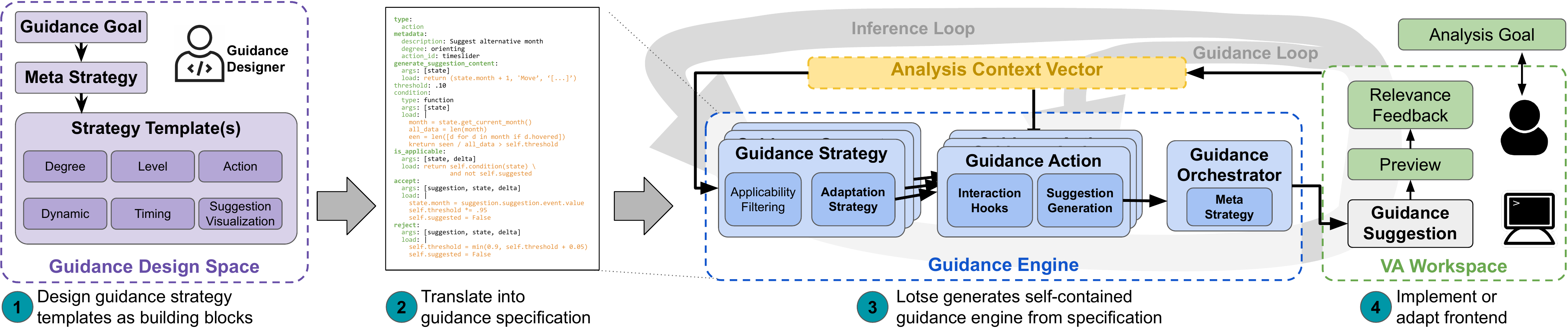}
  \caption{Guidance designers derive strategy templates that encapsulate distinct actions and can be implemented as guidance strategies. Through relevance feedback and temporal adaptation, the engine learns which strategies are helpful in which situations.}
  \label{fig:theory}
}

\vgtcinsertpkg

\begin{document}

\firstsection{Introduction}
\maketitle

While visual analytics (VA) approaches aim to generate insights and knowledge through the promise of effective and efficient visual analysis, actual analysis scenarios often hold potential for time-consuming trial-and-error experiences with increasing frustration. Consequently, research on guidance in VA evolved to become more substantial in recent years\cite{ceneda_characterizing_2017, sperrle_learning_2020, collins_guidance_2018}. Guidance aims to close knowledge gaps that analysts encounter and that keep them from successfully solving a given task. However, previous attempts at characterizing guidance have mostly considered a theoretical perspective where the envisioned benefits remain hypothetical with no directives on applying them in practice. A recent attempt to lay out a design framework for guidance~\cite{ceneda_guide_2020} structures the design process through twelve guiding questions but remains abstract and removed from the actual implementation process. Beyond this work, existing implementations of guidance in VA are mostly bespoke solutions to domain- or task-specific problems with unclear generalizability. 

This paper addresses this shortcoming in guidance research from both a design and implementation perspective. Starting from an analysis of existing guidance approaches and design frameworks, we make a step forward towards their use in practice by developing a practice-oriented model of guidance, shortening the gap between guidance design and implementation. One of the main challenges in implementing guidance systems is developing and selecting guidance actions that are helpful to the user at a given moment. Theoretical guidance frameworks describe processes that first identify when a user needs guidance and then determine which suggestion might help to close their knowledge gap~\cite{ceneda_characterizing_2017, sperrle_learning_2020}. However, this requires guidance designers to rely on the output of intent identification approaches. Despite recent advances~\cite{setlur_data-driven_2020}, intent identification remains a challenging task, and inaccurate results could hamper the success of guidance systems. Instead, we propose an inverted process that is significantly easier to implement. Our model is centered around guidance strategies as the building blocks of adaptive guidance systems. Each strategy should target a specific purpose, such as the  identification of outlier data points, suggestions of alternative model parametrizations, or proposals to use a different visualization (for a range of example strategies, please see \autoref{sec: evaluation}). Developers should implement a wide variety of such strategies and compose them by defining rules describing in which contexts which strategies should be used by the system. Throughout the analysis, these initial rules can then be refined to adapt the provided guidance to the users' requirements. Guidance theory often relies on a top-down approach, in which the system should perfectly model the user and their need for guidance before composing appropriate suggestions. We instead take a bottom-up approach to guidance implementation. We start by asking which suggestions can be implemented in different strategies today and propose that systems should refine when to use which strategy over time. This approach is particularly useful in situations where users perform exploratory analysis and have more than one fixed, well-known analysis goal. 

Building on our practical model of guidance, we introduce \lotse{}, a  library that allows specifying guidance strategies in definition files and converts them to a running guidance system. \lotse{} is inspired by visualization grammars like Vega-Lite~\cite{Satyanarayan_vega-lite_2017} and presents the first step toward a declarative grammar of guidance. \lotse{} represents the guidance process based on the recently introduced \emph{guidance} and \emph{inference} loops~\cite{perez-messina_typology_2022} and monitors the analysis state to determine which strategies to employ and which suggestions to provide. To enable adaptation, \lotse{} offers various hooks allowing developers to customize system behavior on suggestion acceptance or rejection, for example. Using our library, developers are freed from implementing common boilerplate code in guidance orchestration. Instead, they can focus on the design of effective strategies. Our contributions are the following: 
\begin{enumerate}[itemsep=-.4em, topsep=0em]
  \item A summary of theoretical, co-adaptive guidance processes, capturing the current state-of-the-art.
  \item A practical model of guidance focused on guidance templates and strategies to bridge the gap towards system implementation.
  \item A library for guidance specification that generates running guidance systems without requiring writing boilerplate code.
\end{enumerate}

\setlength{\intextsep}{0.5em}
\setlength{\columnsep}{4em}

\section{Related Work}


\paragraph{Visualization Frameworks and Grammars}
There is a vast array of previous work in frameworks for the creation of visualizations. Notably, Bostock et al.~\cite{bostock_d_2011} presented d3, a low-level framework that still seems ubiquitous in visualization development today. Echarts~\cite{li_echarts_2018} takes a slightly different approach to d3 and provides more predefined charts that can be customized. Both libraries require significant implementation work to set up visualizations. To overcome this requirement, various projects from the Vega-universe provide declarative visualization grammars~\cite{satyanarayan_declarative_2014} that even include interactions~\cite{Satyanarayan_vega-lite_2017}. Recently, Gemini~\cite{kim_gemini_2021} defined a grammar for animated visualizations, Cicero~\cite{kim_cicero_2022} facilitates responsive visualizations, and DXR~\cite{sicat_dxr_2019} offers JSON specifications that are parsed into unity prefabs, enabling the declarative specification of immersive visualization components in virtual reality. While these works focus on visualization, \lotse{} shares with them the aim of providing simple yet powerful tools of implementation. In the spectrum between implementation and declaration, \lotse{} offers some declarative features while relying on imperative callbacks and event handlers to realize more complex functionality.

\paragraph{Interaction Grammars}
Vega-Lite~\cite{Satyanarayan_vega-lite_2017}, in addition to offering a visualization grammar, includes a grammar of interaction with said visualizations. Nebula~\cite{chen_nebula_2021} enables the coordination of multi-view interfaces through natural-language specifications following the form ``X then Y''. As such, these rules can also be envisioned to be useful in simple guidance- or onboarding scenarios and could be replicated in \lotse{}. While both Vega-Lite and Nebula, among others (e.g., \cite{satyanarayan_declarative_2014, wongsuphasawat_encodable_2020}) focus on interaction declarations by developers, Dabek et al.~\cite{dabek_grammar-based_2017} analyze user interactions from several dozens of users performing a given, single task to identify frequent interaction patterns. They then use graph algorithms to determine optimal solution strategies from observed user behaviors that they describe using an interaction grammar. If later users of the system deviate from the identified paths, the system aims to guide them back to being more efficient. Gathani et al.~\cite{gathani_grammar-based_2022} automatically extract tasks (according to varying task taxonomies) from interaction logs and encode them as terminals of an interaction grammar. They then identify production rules on those interaction terminals, aiming to decipher the \emph{language} of interaction. \lotse{} provides a first step towards a grammar of guidance, opening a path towards declarative guidance. Because guidance can be described as a continuous process of action-reaction pairs~\cite{sperrle_co-adaptive_2021}, a guidance grammar must, in contrast to existing interaction grammars, cover not only the users' interactions but also interactions initiated by the system to enable contextualized, adaptive guidance. Omitting the system side would mean that only half of the mixed-initiative process could be modeled.

\section{Co-Adaptive Guidance in Visual Analytics}
\label{sec: co-adaptive guidance theory}
Guidance approaches have old roots in human-computer interaction and decision theory~\cite{smith_guidelines_1986,silver_decisional_1991, horvitz_principles_1999,oppermann_adaptive_1994}. Engels et al.~\cite{engels_planning_1996} conceptualized how to provide ``task-oriented user-guidance'' by deconstructing KDD processes into reusable task components. Their theoretical framework is centered around the definition of a goal state for which the system could then propose task compositions leading to this state. This section provides background on recent guidance research in visual analytics. In particular, several theoretical frameworks and characterizations have been proposed in the past few years. As a result, it might be challenging for  novices to identify commonalities between theories of guidance and decide which steps to take when implementing guidance. To support researchers in this task, we summarize existing  theories and characterize guidance along different \emph{guidance levels}.

\subsection{Guidance Theory Review} The most commonly used definition of guidance in VA was coined by Ceneda et al., describing its fundamental characteristics as aiming to \emph{close a knowledge gap}~\cite{ceneda_characterizing_2017} and emphasizing its mixed-initiative nature~\cite{ceneda_guidance_2018}. Their model identifies five domains of knowledge gaps and three guidance degrees to close them. The three guidance degrees \emph{orienting} (highlighting through visual cues), \emph{directing} (ranked list of suggestions), and \emph{prescribing} (system selects \emph{best} suggestion) capture how much control the system takes over the guidance process. Building on that, Sperrle et al. have provided a process model of co-adaptive guidance, highlighting the different guidance dynamics of learning and teaching between humans and machines in this mixed-initiative process~\cite{sperrle_learning_2020}. They state that systems should define clear adaptation goals~\cite{sperrle_co-adaptive_2021} building on Bloom's taxonomy~\cite{bloom_taxonomy_1956} of learning goals.

Collins et al.~\cite{collins_guidance_2018} proposed a process-oriented view on guidance built around a framework of high-level VA tasks~\cite{andrienko_viewing_2018}. They identified several goals of guidance: ``to inform, to mitigate bias, to reduce cognitive load, for training, for engagement, and to verify conclusions''~\cite{collins_guidance_2018} and state that different levels of guidance could exist, ``from low-level operations on adjustment of visual displays to high-level analysis tasks such as model development and evaluation.'' In contrast to Ceneda et al.'s framework that focuses on ``how'' and ``when'' knowledge gaps can be closed, Collins et al. aim to answer ``what'' knowledge users want to derive. However, their provided framework remains high-level and removed from actual implementation practices.

Pérez-Messina et al.~\cite{perez-messina_typology_2022} focus on system tasks in the guidance process and contribute a system task taxonomy that mirrors Brehmer and Munzner's taxonomy of visualization tasks~\cite{brehmer_multi-level_2013}. They further introduce how guidance complements the knowledge generation model~\cite{sacha_knowledge_2014} by including the \emph{guidance loop} and the \emph{inference loop}. Note that both loops are covered by our framework, as shown in~\autoref{fig:theory}.

In addition to these conceptual works, several applications have featured guidance components. For instance, Willet et al. described Scented Widgets that aim to guide the user by integrating hints about other user activities~\cite{willett_scented_2007}. Gotz et al.~\cite{gotz_behavior-driven_2009} provide guidance to users building visualizations by analyzing their interaction history. Sperrle et al. rely on adaptive guidance to support the annotation of arguments~\cite{sperrle_viana_2019} and the refinement of topic models~\cite{sperrle_learning_2021}. An extensive overview of guidance literature is provided in a recent review by Ceneda et al.~\cite{ceneda_review_2019}. While a significant body of work on guidance systems exists, extracting commonalities or general patterns remains challenging. However, developing a ``\emph{better understanding of the internals of guidance}''~\cite{ceneda_characterizing_2017} is necessary as the ``\emph{core function of guidance, i.e., the guidance generation process, largely remains a black box.}''~\cite{ceneda_characterizing_2017}. 

\subsection{Levels of Guidance} 
\setlength{\columnsep}{1em}
From this overview of the guidance literature, we make a further step towards its implementation by abstracting existing  approaches 
\begin{wrapfigure}[12]{r}{0.6\linewidth}
\vspace{-0.5em}
    \includegraphics[width=\linewidth]{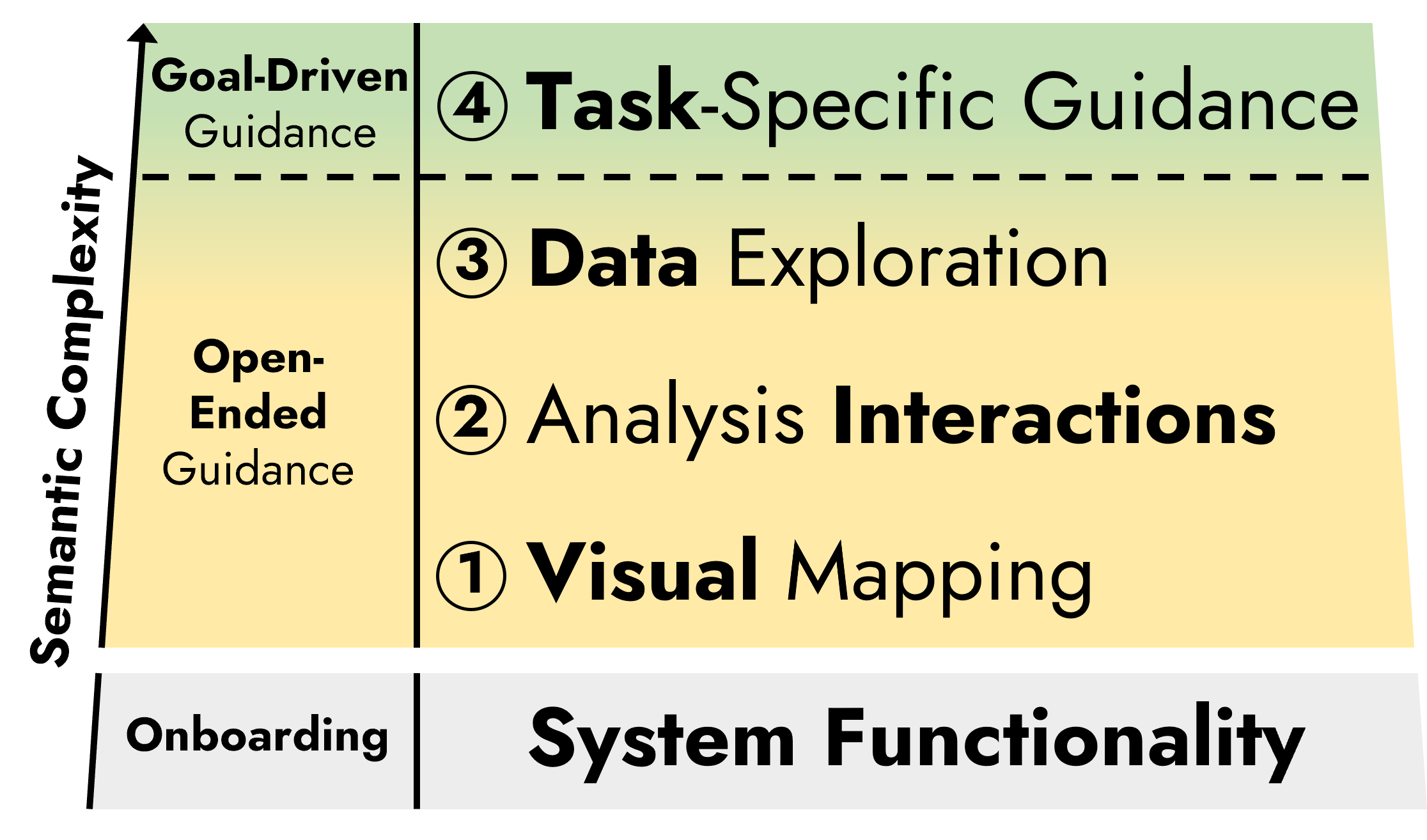}
    \caption{Four guidance levels (in yellow / green) of increasing semantic complexity.}
    \label{fig: pyramid}
\end{wrapfigure}
and reasoning about how systems can provide assistance during VA. We can assume that any VA session starts when a user decides to pursue a given \emph{analysis goal}. To reach their goal and solve their original problem, users typically have to structure it into tasks and complete them sequentially. Along this process, guidance is required whenever the user cannot solve any task on their own. As the  design space is vast, there are many ways in which guidance can be provided. While the overarching aim of guidance should be to close knowledge gaps, their categorization into data, task, and VA methods remains abstract. In \autoref{fig: pyramid}, we introduce four \emph{guidance levels} in two semantic complexity levels \emph{open-ended} and \emph{goal-driven}. In onboarding scenarios, guidance does not apply~\cite{ceneda_guidance_2018, stoiber_perspectives_2022}. The four guidance levels highlight that, from a practical perspective, the implementation of targeted guidance is primarily limited by what is known about the users' tasks and goals: Static tutorials of \emph{system functionality} are \emph{always} possible as \emph{onboarding}. Similarly, guidance on the used \emph{visual mapping} and available \emph{interactions} is \emph{frequently} possible. While \emph{basic} \emph{data exploration} guidance is possible without the notion of a clear analysis goal, more targeted exploration guidance and task-specific guidance are only possible if tasks are clearly identifiable. Hence, the implementation of guidance is constrained by the correct identification of users' intents. It is worth noting that \emph{guidance level} and \emph{guidance degree}~\cite{ceneda_characterizing_2017} are orthogonal concepts. The guidance level describes the semantic complexity of a given suggestion, while the guidance degree characterizes how much control the system assumes over the mixed-initiative analysis process.

Some approaches try to identify the user intent and then provide suitable guidance (e.g., \cite{gotz_behavior-driven_2009}). However, detecting intents can be challenging, especially if no situational knowledge base or analytical provenance data can be exploited. Hence, other approaches offer \emph{speculative guidance} to probe the user for potential knowledge gaps and gather relevance feedback~\cite{sperrle_speculative_2018}. This process continues  until the task (and possible knowledge gaps) is identified, or enough relevance feedback is gathered. Then, the system is ready to provide guidance. Our model of strategy-based guidance that we present in this paper follows a similar approach.

Generating guidance subsumes the consideration of the analysis state but also of the type of analysis carried out, i.e., open-ended vs. task-driven analysis. Designers have an active role in deciding how the system could respond to certain events by specifying rules and defining guidance goals. In summary, we see the generation of guidance as the result of two processes: (1) The definition of \textit{static} directives specified at design time to tackle \emph{expected} adverse events, and (2) how these rules translate into practice, given the \textit{dynamic} nature of the analysis and the settings of the scenario. In the next section, we refer to these aspects as guidance templates and guidance strategies and show how they can be used to implement guidance in practice. 

\section{A Strategy-Centered Guidance Process Model}
\label{sec: guidance model}
\setlength{\columnsep}{1em}
Taking the constraints discussed above into account, \autoref{fig:theory} illustrates a suggested way to design and implement contextualized, adaptive guidance. Our implementation model is primarily targeted at  problems that are situated at the transition between open-ended and goal-driven guidance. More specifically, we envision our model to be useful when user tasks contain a mixture of open-ended data exploration and specific tasks on previously identified data points. The two core components of our approach are \emph{strategy templates} as instances of the guidance design space and the corresponding \emph{guidance strategies} that provide concrete implementations on the system level. In this section, we present the theoretical model of strategy-based guidance. In \autoref{sec: lotse} we then introduce \lotse{}, our  library that instantiates this model.

Designing guidance begins with an exchange between designers and potential users to determine the goals of the guidance and key characteristics like \emph{degree}, \emph{level}, \emph{dynamics}, and \emph{timings} that will be introduced in detail below. This theoretical design process has been characterized in terms of twelve guiding questions by Ceneda et al.~\cite{ceneda_guide_2020}. A common problem of theory-driven guidance is the focus on systems inferring the users' knowledge gaps before providing suggestions. As this is typically challenging, we propose to invert this process and begin by designing possible guidance strategies. Once potential strategies are defined, rules for employing those strategies can be devised and then refined during the analysis session. By outsourcing some challenges to the  designer, we provide an actionable mapping from their design decisions (the templates) to implementable  strategies. Each strategy template covers specific \emph{actions} that can be turned into suggestions. 

\begin{wrapfigure}[18]{r}{0.30\columnwidth}
\vspace{-0.2em}
\includegraphics[width=\linewidth]{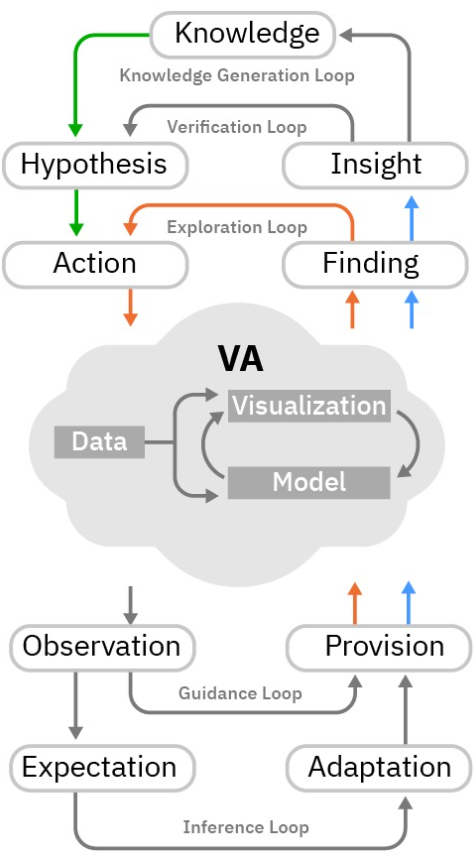}
\caption{Guided knowledge generation model~\cite{perez-messina_typology_2022}. 
}
\label{sec: guided knowledge generation model}
\end{wrapfigure}

The resulting guidance process envisions two main loops between the user and the system or, more specifically, the guidance engine of the system. These loops correspond loosely to the \emph{guidance loop} and \emph{inference loop} of the guided knowledge generation model (see \autoref{sec: guided knowledge generation model}) that we introduced in previous work~\cite{perez-messina_typology_2022}. In the outer loop, the system determines which guidance strategies are applicable in the 
current analysis state based on what it expects the user's task to be. Each applicable strategy defines \emph{conditional actions} that are evaluated and ranked by the \emph{guidance orchestrator} component before being presented to the user as  suggestions. In the inner loop, the system evaluates obtained relevance feedback to previous suggestions together with the progressing analysis to re-evaluate the available actions. Whenever the focus of the analysis changes, e.g., because the user has reached their goal, the outer loop enables the selection of other, more appropriate strategies, whereas the inner loop focuses on providing appropriate suggestions from the currently used strategies and their adaptation.

In the following, we describe the components of the proposed model in an order aligned to the design process. Using the color map from \autoref{fig:theory}, we introduce each  component with a box containing a short definition and definitions of relevant key terms, where appropriate. 

\subsection{Strategy-Centered Guidance Design}
\label{sec: strategy-centered guidance design}
Similar to the VA analysis process, guidance design begins with eliciting user requirements, goals, and potential knowledge gaps~\cite{ceneda_guide_2020}. As these intents can often not be identified with precision (e.g., in open-ended analysis) or users switch between different tasks and objectives, we rely on designers providing a variety of strategies and context rules for when to use them. Over time, the guidance system can then adapt those initial rules based on observations from the analysis process, e.g., in the form of relevance feedback. Throughout the section, we rely on a running example to illustrate the components of our model. Our envisioned example system offers a scatter plot showing weather data of cities all over the world and a time slider that allows users to scroll through different months in the scatter plot. Users can map different dimensions (temperature, precipitation, wind, etc.) to both axes of the scatter plot. Assume the users' goal is to explore the data to find a vacation time and target. To assist them, the guidance system encourages exploration of all months and highlights data points of particular interest. 

\component{Analysis Goal}{user}{
A description of the goal the users want to reach during analysis. Similar to Gotz et al.'s~\cite{gotz_behavior-driven_2009} multi-level task classification (event, action, sub-task, task), goals exist in a hierarchy of increasing complexity.
}
\noindent In visual analysis, users have one or multiple goals. As outlined above, various theoretical guidance frameworks are centered around the notion of identifying knowledge gaps or what knowledge the user aims to derive. Correctly capturing these goals--in our example, the identification of a month and a city--and the underlying intents is vital for generating effective guidance. However, in practice, many guidance approaches support specific tasks rather than open-ended exploration and avoid intent identification due to its inherent difficulties. In that case, the success of the provided guidance relies on the requirements elicited in the design process. Once the main user intents have been identified, the guidance goals in the analysis should be identified.

\component{Guidance Goal}{designer}{
The guidance goal describes the high-level intent of the guidance system and abstracts the underlying guidance actions.
\vspace{-0.5em}
\begin{description}[wide]
    \item[Guidance Dynamic] Characterizes the interaction dynamics between user and system from the perspective of the respective intents~\cite{sperrle_learning_2020}.
\end{description}
}

\noindent Among others, we recognize two high-level guidance goals: \emph{support} and \emph{correct}. The \emph{support} goal captures systems identifying that the user is on the correct path to reaching a goal and providing suggestions to make them, for example, faster. In that sense, it does not aim to change the user's goal but might change the user task needed to reach said goal. In contrast, the \emph{correct} goal is aimed at getting the user to change their analysis behavior. As a result, interrupting them in their analysis is not only acceptable but desired. Defining an appropriate guidance goal is crucial to successfully implementing guidance. As goals are associated with different interaction dynamics between user and system, the defined guidance goal shapes the analytical discourse between system and user. In the weather analysis example, we chose supporting guidance, as there is no information based on which the system could determine that it would need to correct user behavior.

\component{Strategy Template}{designer}{A strategy template is a summary of an envisioned guidance strategy containing all information necessary to implement the strategy.
\vspace{-0.5em}
\begin{description}[wide]\setlength\itemsep{-0.3em}
    \item[Guidance Degree] Orienting, Directing, Prescribing~\cite{ceneda_characterizing_2017}.
    \item[Guidance Level] Guidance levels (see \autoref{fig: pyramid}) describe the semantic complexity of suggestions.
    \item[Timing] The  timing determines in which situations a template should be used. Depending on the system, this decision might happen on different granularities -- either as broad analysis contexts in which a suggestion might be useful or as concrete measures (e.g., ten seconds after the mouse stopped moving).
    \item[Action] The concrete content or impact of a suggestion. 
\end{description}}

\noindent Guidance templates serve as design sheets in the process of developing strategy-based guidance and collect all information necessary for later implementation. In our model, each  template corresponds to exactly one  strategy that will be implemented later and thus corresponds to exactly one type of suggestion that could be shown to the user during the analysis--for example, switching to a different month or considering specific data points. This strict limitation introduces a separation of concerns between strategies, limiting their complexity with the goal of simplifying their implementation. We believe that this separation is particularly useful for novice guidance designers. For them, thinking about individual strategies at a time provides clear steps in the design process that translate into individual implementation tasks later. Designers should consider different types of strategies, e.g., branching out, reinforcing, or serendipity~\cite{collins_guidance_2018}. Within each type, multiple strategy templates can be envisioned that offer distinct suggestions.

\begin{table}[b]
\begingroup
    \vspace{-1em}
    \centering
    \renewcommand{\arraystretch}{1.02}
    \begin{tabular}{| >{\columncolor{gray!20}}p{0.1\columnwidth}p{0.82\columnwidth}|}
    \toprule
    Basic & \textbf{Strategy Name}:  \hrulefill\\
    Info & \textbf{Description}:  \hrulefill\\
    & \hrulefill\\
    
        \hline 
        
    Goal & \textbf{Guidance Goal}:  {\small\faCircleO}~Support \quad {\small\faCircleO}~Correct\\
    & \textbf{Level}:  {\small\faCircleO}~Onboarding \quad {\small\faCircleO}~Open-Ended \quad {\small\faCircleO}~Goal-Driven\\
    & \textbf{Dynamic}: {\small\faCircleO}~Learn \quad {\small\faCircleO}~Teach \quad {\small\faCircleO}~Other~\hrulefill \\
    & \textbf{Knowledge Gap Type}: {\small\faSquareO~Path to Solution}~{\small\faSquareO~Optimal Solution}\\
    & \textbf{Knowledge Gap Domain}: {\small\faSquareO}~Data~{\small\faSquareO}~Tasks~{\small\faSquareO}~VA Methods\\
    & \textbf{Details}: \hrulefill\\
    
        \hline 
        
    Action & \textbf{Content}:  \hrulefill\\
    & \hrulefill \\
    & \textbf{Timing}: {\small\faCircleO}~Static ~{\small\faCircleO}~Contextualized on: \hrulefill\\
    & \textbf{Adaptation}: {\small\faCircleO}~No \quad {\small\faCircleO}~Yes: \hrulefill\\
    
        \hline 
        
    Visual & \textbf{Degree}:  {\small\faCircleO}~Orienting \quad {\small\faCircleO}~Directing \quad {\small\faCircleO}~Prescribing\\
    ization & \textbf{Placement}:  {\small\faCircleO}~In-situ \quad {\small\faCircleO}~Ex-situ \\ 
    & \textbf{Visual Elements}: \hrulefill\\

    \bottomrule
    \end{tabular}
    \endgroup
    \caption{A generic example guidance strategy template form. Depending on the envisioned guidance, more concrete fields should be added. }
    \label{tab:template}
    
\end{table}

We provide a template form that  designers can fill out during the design process in~\autoref{tab:template}. In addition to the intended  action, templates capture various theoretical aspects of guidance like  degree, dynamics, level, and timing, but also the visual representation of suggestions. Consequently, the  template generation is the  designers' main task, as a guidance template is a blueprint for implementing a guidance strategy, similar to specification sheets typically used in software development.

\component{Meta-Strategy}{designer}{The meta-strategy mediates between the user's intent and the most relevant suggestions identified by the guidance system.}

\noindent Meta-strategies allow designers to define \textbf{balancing-mechanisms} for prioritizing and alternating the different guidance strategies. Through the dynamics defined in the guidance goal, a meta-strategy encodes how aggressively the user is led onto a path suggested by the strategies. In addition, it provides information on how to select the next guidance action(s) to be suggested in ambiguous cases. Simple meta-strategies include, for example, selecting actions that are (dis)similar to previously suggested actions or from the last user actions. In our example, we prioritize finishing the exploration of a selected month over potential suggestions to switch months. Guidance is typically concerned with modeling the user's knowledge and identifying gaps therein. In that sense, the combination of guidance goal and meta-strategy can be seen as an expression of the designer's previous knowledge about the analysis process, complementing the user's knowledge.

\component{Relevance Feedback}{user}{
Provided by the user to the system, either explicitly or implicitly.
\vspace{-0.5em}
\begin{description}[wide]\setlength\itemsep{-0.3em}
    \item[Explicit Feedback] Is provided with the intent of giving feedback.
    \item[Implicit Feedback] Is automatically derived from the user's actions.
\end{description}}

\noindent To enable dynamic system behavior, guidance designers need to consider whether and to what extent they would like to incorporate user feedback into the generation of  suggestions. Next, they need to determine whether (and how) it is feasible to adapt the content of guidance suggestions or whether the focus is on adapting the timing or other properties like degrees. Both aspects are captured in the strategy template under \emph{timing} and \emph{adaptation}. In general, systems can process explicit user feedback and automatically infer implicit feedback from their interactions (\emph{backward feedback}). Alternatively, users can signal to the system what they expect as future guidance (\emph{forward feedback})~\cite{ceneda_review_2019}. Once collected, the feedback serves to steer the selection of strategies and actions. 

\subsection{Modeling Users, Data, and Tasks}
Ideally, guidance is dynamically generated, matching the designer's directives to the specific necessities of the user and analysis state. 

\component{Analysis State}{knowledge}{We define the analysis state as all collected information necessary to determine which strategies and actions to employ.}

\noindent The analysis state is at the center of each VA process and captures any information that is relevant for either the analysis or future guidance. In the guidance template in \autoref{tab:template} it can be found under \emph{action timing}. We omit definitions and descriptions of the well-established task, data, and user models here and refer readers to previous work~\cite{sperrle_learning_2020}.  Practical implementations of guidance can rely on a state model that serves as a catch-all model for relevant information, such as application settings and previous  suggestions, but also analyzed machine learning models, visualization snapshots, and the gathered relevance feedback mentioned earlier. In our example, we store a model of the data itself, the dimensions shown in the scatter plot, the current month, and which data points the user interacted with already. In theory, many distinct snapshots of the analysis state-space could be maintained over time to capture the evolving analysis. However, such extensive data capture is expensive in terms of computing power and storage. As a compromise, storing the \emph{delta} to the previous state vector can already provide meaningful added information. 

\subsection{System Components of Strategy-Based Guidance}
While \autoref{sec: strategy-centered guidance design} focused on a strategy-centered guidance design process, this section introduces the resulting system components of strategy-based guidance. A specific implementation of these components in \lotse{} will be introduced in \autoref{sec: lotse}, together with further examples. 
    
\component{Guidance Strategy}{system}{
Guidance strategies are concrete implementations of strategy templates and are responsible for generating guidance actions. 
}
\noindent Strategy templates devised in the design phase can each be implemented into a strategy when developing the VA system. The more carefully the  templates were designed, the more straightforward their implementation becomes. As Collins et al. stated, guidance ``should be adapted to the context of the analysis process''~\cite{collins_guidance_2018} as ``different kinds of guidance could be reasonable''~\cite{collins_guidance_2018} in different phases. Strategy-based guidance tackles this adaptation from two perspectives. First, the availability of distinct strategies and analysis-state-based rules on when to employ them lay the foundation for adaptive guidance. To that end,  strategies contain an \textbf{applicability filter} that defines in which broad analysis contexts the strategy should be available and is evaluated by the \emph{inference loop}. Second, both strategies and actions offer adaptation strategies that process implicit or explicit user feedback into new application rules. This dual approach enables tailoring contextualized guidance to a specific user over time. 
    
\component{Guidance Action}{system}{Guidance actions belong to a strategy and describe how the content of suggestions from that strategy is produced, how suggestions are timed, and how they are adapted. }
\noindent Guidance actions encapsulate the \textbf{generation of suggestions} based on the definition in the  template. While  strategies determine the broad contexts in which they should be available, individual actions offer more precise filter rules that are continuously re-evaluated by the \emph{guidance loop}. If an action is applicable in a specific analysis state, it generates a new suggestion according to the definition in the  template. Our model does not impose restrictions on the types of suggestions. Common examples of suggestions include encouraging users to switch to different views, considering alternative data, or changing algorithm parameters. With \textbf{interaction hooks},  actions also provide a place in which the  system can react to captured relevance feedback and update the applicability rules of both actions and strategies.  

\component{Guidance Orchestrator}{system}{The guidance orchestrator is the technical instantiation of the meta-strategy devised in the design phase.}
\noindent The guidance orchestrator is responsible for prioritizing  applicable  actions in a given context. When multiple actions are deemed to be applicable, they are passed to the guidance orchestrator. The orchestrator filters or prioritizes actions based on the defined meta-strategy, acting as a balancing mechanism between competing strategies. Potential criteria for filtering could include a comparison of provided model quality improvements, time saved, and envisioned guidance dynamics or the (dis)similarity from previous suggestions. Ultimately, all suggestions that the  orchestrator does not filter out are turned into  user suggestions. 

\section{The \lotse{} Guidance Library}
\label{sec: lotse}
\autoref{sec: guidance model} focused on the introduction of strategy templates as elements supporting a structured guidance (design) process and introduced the concept of strategy-based guidance. In this section, we present \lotse{} (German for \emph{guide}), a  library allowing guidance developers to instantiate their created strategy templates via \texttt{yaml} files, similar to visualization grammars such as Vega-Lite~\cite{Satyanarayan_vega-lite_2017}. The created  strategies and the analysis state definition are then automatically compiled to a stand-alone guidance system ready to provide suggestions. \lotse{} is implemented in Python (data handling, guidance generation) and available at \url{https://pypi.org/project/Lotse/} or at \url{https://github.com/lotse-guidance}, where we offer implementation examples and visualization frontend adapters for d3~\cite{bostock_d_2011}, Vega~\cite{Satyanarayan_vega-lite_2017}, \emph{angular} and \emph{react}. Storing data into \lotse's state does not depend on those frameworks, though, and can easily be added to any system that can perform REST requests. Rather than writing yaml files, experienced \lotse{} developers can also directly implement \lotse{}'s internal python classes.  

\paragraph{Design Goals}
When designing \lotse{}, we aimed to produce a powerful, generic way to implement guidance efficiently. In particular, we adhered to the following design goals that capture theoretical requirements of guidance (mixed-initiative and adaptive) and are relevant to novices (easy-to-use strategies): 

\noindent\textbf{Ease of Use:} First and foremost, we aimed to make \lotse{} easy to use to reduce the barrier of entry into guidance design for VA experts that might question where to begin implementing guidance today. Hence, \lotse{} provides defaults that allow developers to drop in strategies and actions and immediately observe initial  suggestions.

\noindent\textbf{Strategy-Based:}  Reasoning about individual  strategies and how they fit together to reach a guidance goal provides a structured way to design guidance--the same benefits apply during implementation.

\noindent\textbf{Mixed-Initiative:} \lotse{} features an internal event loop that is used to schedule guidance operations, allowing it to produce suggestions without the need for explicit requests. To transport generated suggestions to the visualization frontend, \lotse{} thus relies on websockets to distribute new  suggestions (in contrast to, e.g., forcing developers to implement pulling operations via REST interfaces). 

\noindent\textbf{Adaptive:} Which suggestion is best suited for a given user at a given point in time might be difficult to premeditate when designing guidance. As a result, \lotse{} offers various suggestion interaction hooks, allowing developers to define callbacks and initiate adaptation.

\paragraph{Architecture Summary}

\lotse{} is centered around the concept of \textbf{guidance strategies} introduced above. Each strategy defines \textbf{conditions} based on the \textbf{analysis state} in which its \textbf{guidance action} should be triggered, leading to a new \textbf{suggestion}. Suggestions, in turn, define \textbf{interaction hooks} that can be used to drive adaptation. As the analysis progresses, changes to the analysis state are recorded as \textbf{state updates}. \textbf{Strategy orchestration} allows designers to enable or disable strategies in given conditions.  Both analysis states and guidance strategies can be specified in yaml files that \lotse{} parses into running python code at startup. In several locations, developers can specify hooks or callbacks to define the concrete behavior of the created  system. In the remainder of this section, we present those key concepts in detail, outlining \lotse{}'s architecture. We continue to use the example from \autoref{sec: guidance model}. The two strategies highlight outliers in the scatter plot as they might be of particular interest and suggest moving to a different month when enough data points have been explored, respectively. This section introduces the corresponding strategies, trigger rules, and adaptations. All presented code is available on GitHub.

\paragraph{\lotse{} Grammar} \lotse{} allows specifying both analysis states and (meta) strategies as yaml files. We rely on a very relaxed grammar for this purpose. To define custom callbacks, developers specify \yaml{type: function}, an optional list of arguments \yaml{args: []}, and the callback under \yaml{load: ...}. Individual, reserved keywords (e.g., data loading, predefined interaction hooks, etc.) will be introduced below. Beyond those reserved keywords, all other properties (e.g., maps, lists, strings, etc.) specified in the yaml files will be parsed and added as fields into the respective entities. From there, they are available for use in all callback of their respective entity using python-like syntax under \yaml{self}. This freedom enables developers to specify strategies, actions, and state vectors of higher complexity, where necessary. 

\subsection{Analysis State Representation}
\begin{figure}[b!]
    \vspace{-1em}
    \centering
    \includegraphics[width=\linewidth]{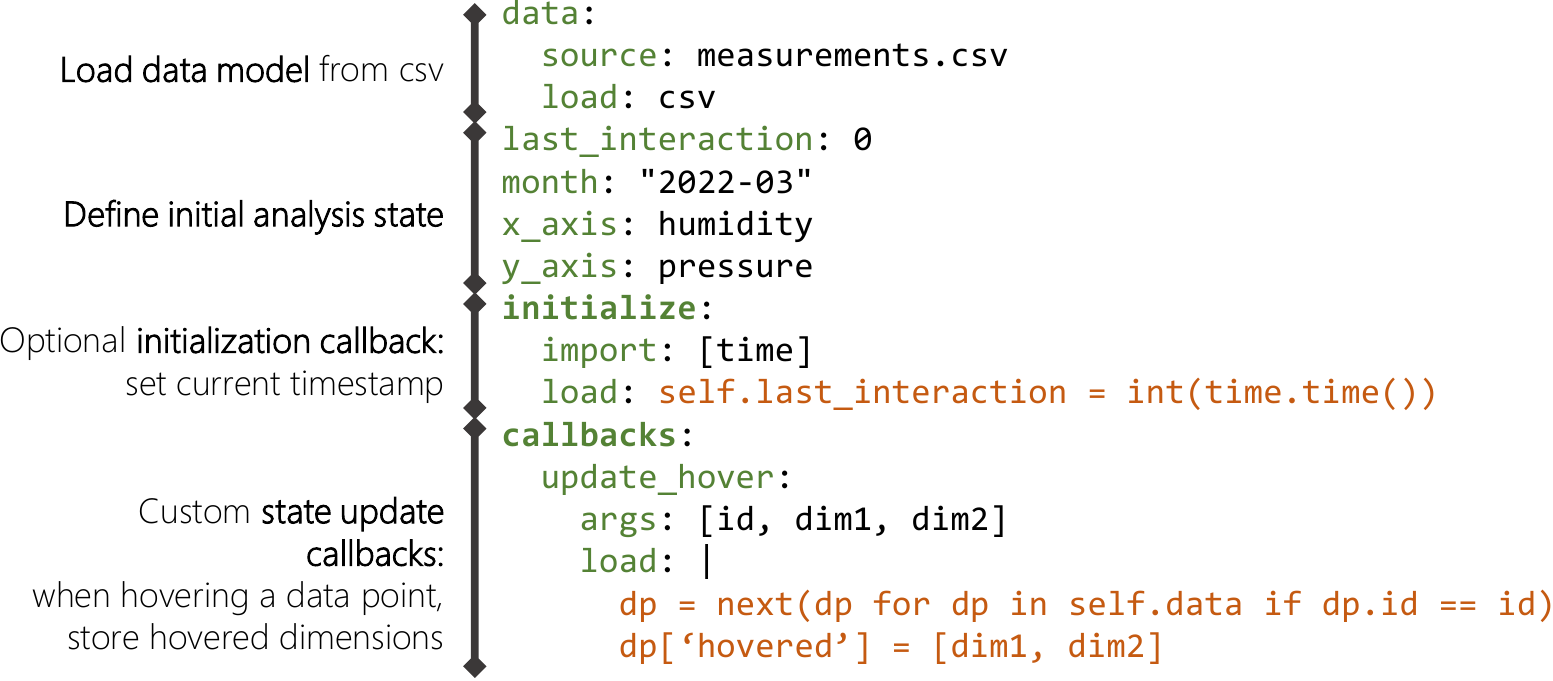}
    \caption{An analysis state definition in \lotse{} defines relevant properties as well as initialization- and update-callbacks.}
    \label{lst:context vector}
\end{figure}

\lotse's core data structure is the analysis state vector. This state vector holds a bespoke representation of the analysis state that is customized for each guidance application. Typical state vectors include representations of the data under analysis, user interaction (meta)data, or model quality metrics. More advanced implementations could also integrate user preference models or modeling alternatives. \lotse{} does not impose restrictions on the amount or type of data stored in the state vector.

Analysis state vectors, like the guidance strategies introduced below, are defined as yaml files to reduce coding efforts. An example state vector is shown in \autoref{lst:context vector} and consists of three sections. In the first section, \lotse{} allows defining arbitrary values into the state vector. All definitions follow default yaml syntax, e.g., \yaml{month: "2022-03"}. \lotse{} recognizes the key \yaml{data}: to load any data from a file or URL, specify the sub-key \yaml{source} and \yaml{load: csv} or \yaml{load: url}, respectively.

\paragraph{Initializing Analysis States}
In some cases, initializing the analysis state with real-time information or complex data structures might be necessary when starting \lotse{}. For that purpose, an \yaml{initialize} callback can be overridden in the analysis state definition. In the example in \autoref{lst:context vector}, this callback is used to store the system startup time into the state. As all callback definitions in \lotse{}, the callback code is defined in \yaml{initialize.load}. All attributes from the analysis state (in this example, the data) are available in python-like syntax under \python{self} leaving developers with full flexibility. 

\paragraph{Updating the Analysis State}
\lotse{} offers two methods for (partially) updating the state vector with corresponding REST interfaces and callbacks defined in the state vector. For simple updates, \lotse{} supports \textbf{setting properties} by sending a JSON object containing key-value pairs to be set on the state vector. More complex updates can be realized via \textbf{callbacks} defined in the analysis state. Developers can define as many callbacks as necessary. In the example in \autoref{lst:context vector}, we use a callback to update our data model to store that a data point has been hovered. Callbacks require specifying a list of expected arguments and a code snippet. Both context update methods are exposed via REST API endpoints. As \lotse{} is independent of any (visualization) framework, developers are responsible for integrating calls to the API. Such calls could be made from the frontend (e.g., when users perform specific interactions) or an existing backend (e.g., when complex data analysis returns new results). Example integrations are available on GitHub.

\subsection{Guidance Strategies and Orchestration}

\begin{figure}[b]
    \vspace{-1em}
    \centering
    \includegraphics[width=0.9\linewidth]{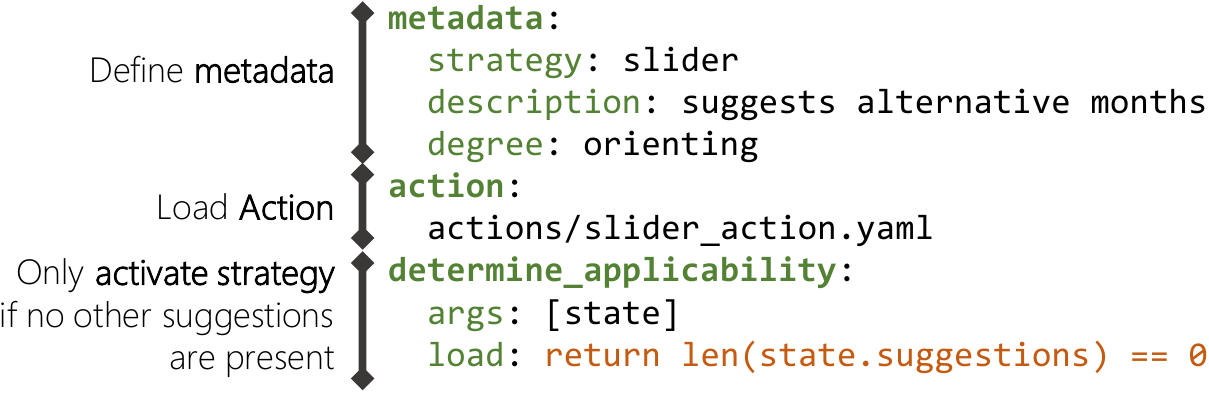}
    \caption{A strategy example in \lotse{} references an action and defines an applicability callback.}
    \label{lst:strategy}
\end{figure}
Guidance strategies are concrete instantiations of strategy templates and can be defined as yaml files (see \autoref{lst:strategy}). They define methods to \textit{determine the applicability} of the strategy and the available \textit{actions}, as well as \yaml{metadata} required for the visualization, such as a \yaml{strategy} name and optional \yaml{strategy_id}, a guidance \yaml{degree} and a human-readable \yaml{description}. The guidance action is stored in its own yaml file and referenced under \yaml{action}. Actions will be introduced in the following section. To orchestrate the guidance process, \lotse{} runs both the inference loop (see \autoref{sec: guidance model}) and the guidance loop on configurable interval timers. In the outer \emph{inference loop}, the system determines which strategies are currently active and should potentially produce suggestions. To enable this functionality, each strategy must define a \yaml{determine_applicability} callback which takes the analysis state as a single argument and returns \python{True} or \python{False}. Activating and deactivating strategies in the inference loop allows developers to change the broad strokes of the provided guidance. 

\subsection{Guidance Actions and Suggestions}

\begin{figure}[t]
    \centering
    \includegraphics[width=\linewidth]{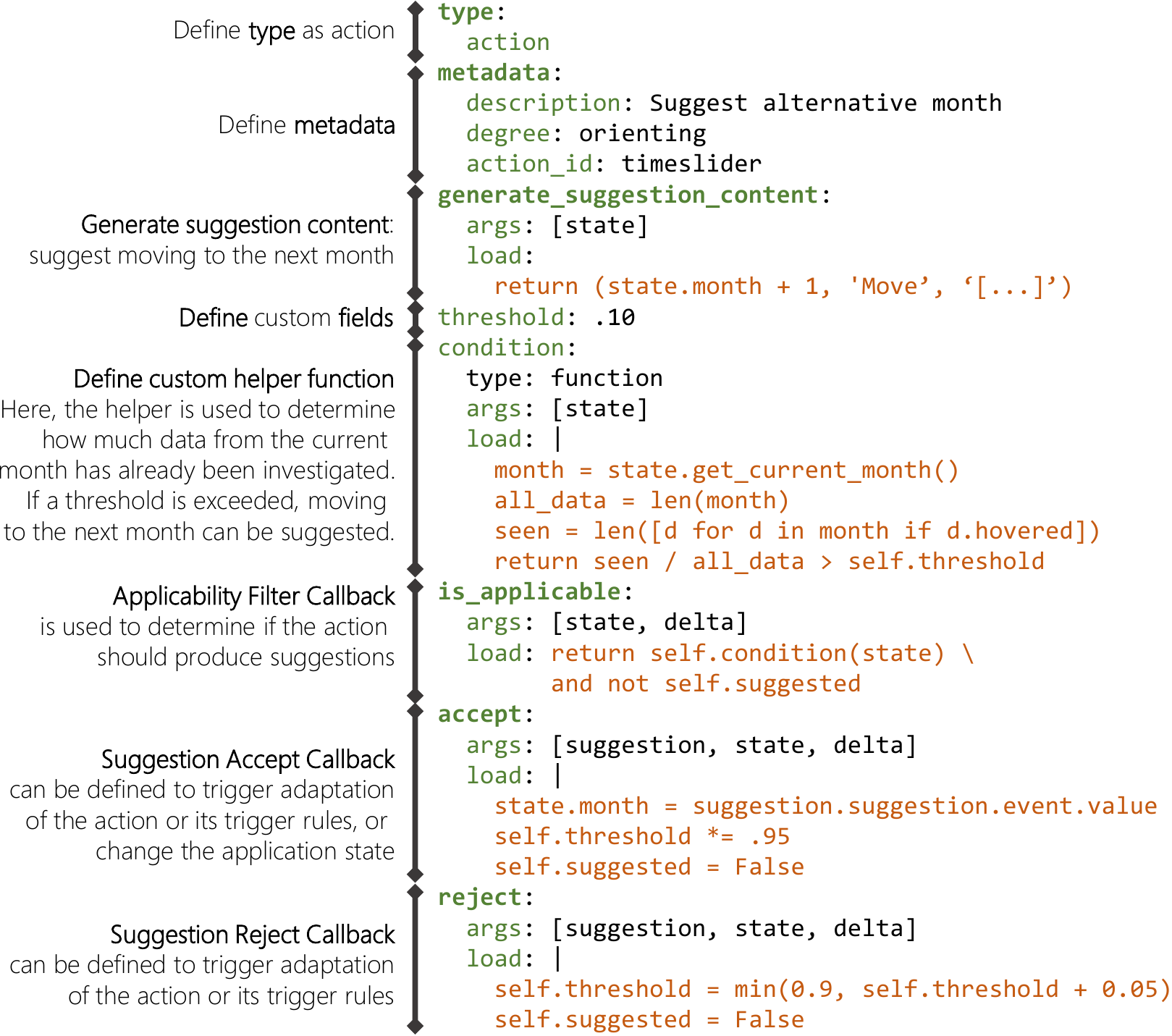}
    \caption{An action in \lotse{} defines how suggestion content is generated, how the action's applicability is determined, and what happens when suggestions are accepted, rejected, or previewed (not shown). }
    \label{lst:action}
    \vspace{-1em}
\end{figure}

Guidance actions are also defined in yaml files, together with some \yaml{metadata}. In particular, each action must define a \yaml{action_id} that will be added to all suggestions it produces. 

\paragraph{Suggestion Generation and Retraction}
In the inner \emph{guidance loop}, the system evaluates conditions defined by each strategy's action to determine whether to generate a new  suggestion or not. By default, the guidance loop is triggered every second, and the inference loop every 30 seconds. Both parameters can be adjusted to fit specific implementation needs. The guidance loop immediately restarts whenever the analysis state is updated. 

Each action must define rules (on the analysis state) for when it is ready to generate suggestions. To that end, the \yaml{is_applicable} callback is used and receives the current analysis state as an argument. If this callback returns \python{True}, \lotse{} generates a new suggestion by calling \yaml{generate_suggestion_content}. This callback must return the suggestion content and a title and description of the suggestion to encourage explainable guidance. The generated suggestions can contain arbitrary data, as long as it can be serialized to JSON to be transmitted to the user interface, leaving developers complete freedom over how to design their guidance. In the example in \autoref{lst:action}, we simply suggest the next month, but more complex suggestions such as computing a DBSCAN~\cite{ester_density-based_1996} clustering on the data points and suggesting outliers are possible (see GitHub for examples).

In addition to generating new suggestions, the guidance loop also verifies whether previously made suggestions have become outdated. To that end, all actions can define \yaml{should_retract} callbacks that receive a context-suggestion pair and return \python{True} if the suggestion should be retracted. Suggestion retractions are also sent via the websocket. 

\paragraph{Meta-Strategy: Suggestion Selection}
If multiple actions produce suggestions at the same time, the meta-strategy is provided with all produced suggestions to its \yaml{filter_suggestions} callback. The meta-strategy can then decide which suggestion(s) should actually be made, e.g., by considering similarity to previous suggestions. 

\paragraph{Suggestion Transmission} 
\begin{wrapfigure}[11]{r}{0.5\columnwidth}
    \includegraphics[width=\linewidth]{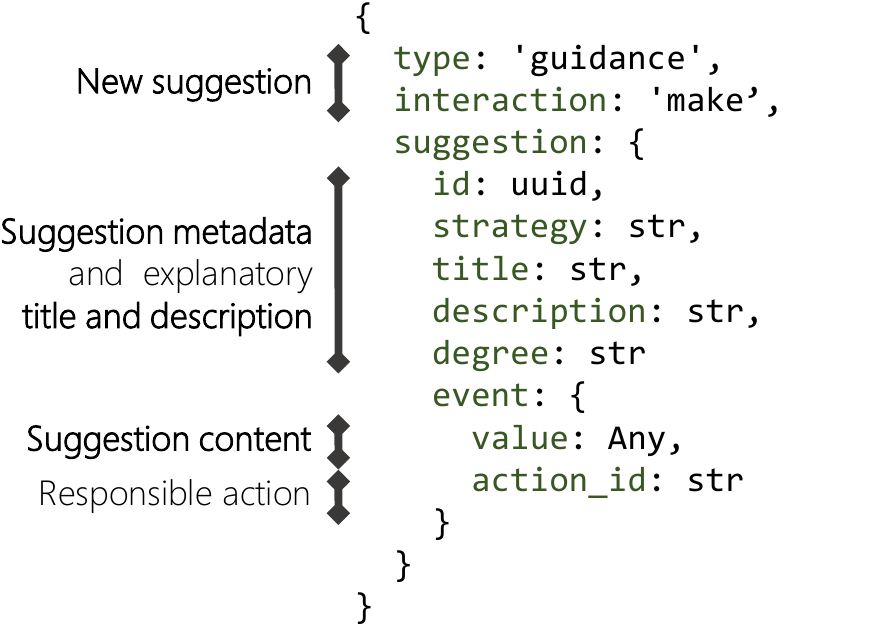}
    \caption{Suggestion data structure}
    \label{lst:suggestion}
\end{wrapfigure}
Once generated,  suggestions are provided via a websocket that the frontend application can subscribe to. Using a socket allows asynchronous guidance updates and enables the system to take the initiative without having to wait for a frontend request. In their frontend, developers subscribe to guidance messages from the socket. For JS-based applications, we offer example configurations for angular, react, and plain javascript.

Before transmission, suggestions are serialized to JSON. We provide an example of the resulting structure in \autoref{lst:suggestion}. Each suggestion gets a unique ID to simplify future communication and contains the names of the strategy and action that were used to generate it. The suggestion content can be an arbitrary JSON object. To encourage the development of explainable and justifiable guidance, \lotse{} suggestions contain a \yaml{title} and \yaml{description} field that developers can supply accordingly. 

\paragraph{Interaction Hooks}
\lotse{} provides four interaction states for each suggestion via REST endpoints. All endpoints expect as payload the guidance message to interact with. Most importantly, users can \textbf{accept} or \textbf{reject} suggestions. To capture more detailed interaction data, additional endpoints for \textbf{starting} and \textbf{ending previews} are also available. All endpoints automatically call the corresponding \yaml{accept}, \yaml{reject}, \yaml{preview_start} or \yaml{preview_end} callbacks defined in the action's specification file. Developers can optionally implement those callbacks to perform additional actions, such as updating when actions should generate suggestions based on the relevance feedback received.

\paragraph{Adaptation}
As outlined above, the success of quality-based strategy implementations depends largely on systems being able to adapt in which contexts they present which strategies. In \lotse{}, adaptation is driven via rule-adjustments executed from the \yaml{accept}, \yaml{reject}, or \yaml{preview} callbacks. Developers can then adjust trigger thresholds, update rule sets, or re-train ML models as needed from those callbacks. 

\section{Evaluation}
\label{sec: evaluation}
We rely on multiple evaluations to demonstrate the applicability and usability of \lotse{}. First, we present usage scenarios showcasing common guidance issues and replicating existing systems. To evaluate the usability of \lotse{}, we conducted a first-use case study with VA experts. We use the terms \emph{usage scenario} and \emph{case study} as suggested by Sedlmair et al.~\cite{sedlmair_design_2012}. We also provide an evaluation of \lotse{} in the \emph{cognitive dimensions of notation}~\cite{blackwell_cognitive_2001} as supplemental material.

\subsection{Usage Scenarios}
We present two usage scenarios for strategy-based guidance using \lotse{} that replicate existing guidance approaches or showcase opportunities. The scenarios focus on 1) providing suggestions contextualized on user behavior and 2) adapting suggestions via relevance feedback. A third scenario replicating Scented Widgets~\cite{willett_scented_2007} and showcasing the real-time capabilities of \lotse{} can be found in the supplementary material. The scenarios demonstrate that strategy-based guidance can be used in a variety of situations and should serve as an inspiration for how to use \lotse{} in VA systems, not as an exhaustive list of what is possible with it. Hence, we describe all scenarios from the perspective of a guidance designer.

\subsubsection{Behavior-Driven Visualization Recommendation}
\begin{figure}[b]
    \centering
    \vspace{-1em}
    \includegraphics[width=\linewidth]{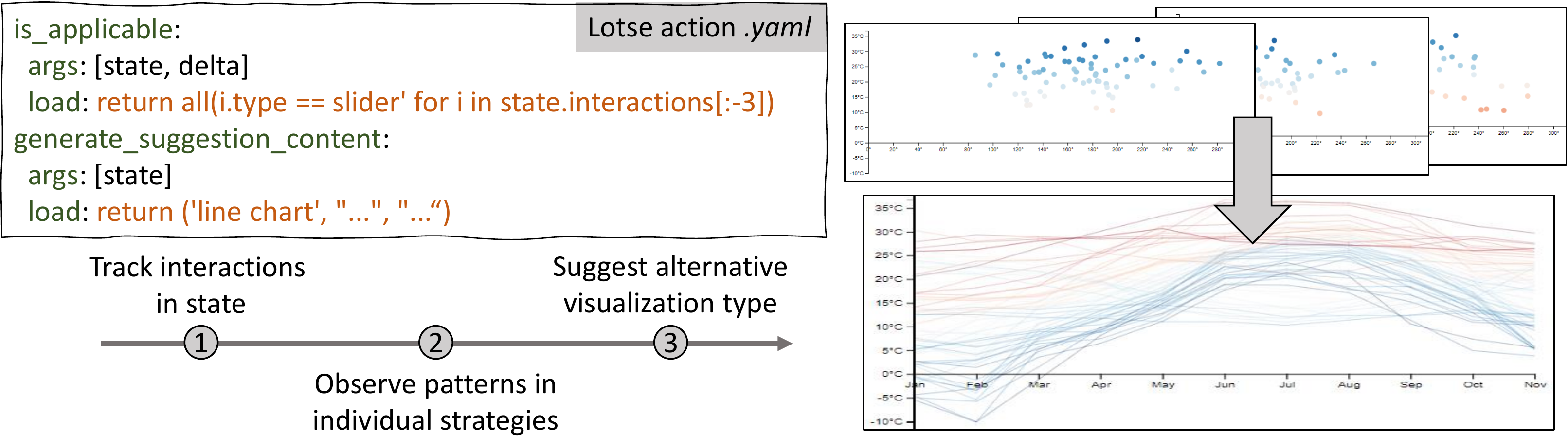}
    \caption{\lotse{} can be used to react to streams of interaction patterns, e.g., to recommend alternative visualizations. }
    \label{fig:patterns}
\end{figure}

Inspired by Behavior Driven Visualization Recommendation~\cite{gotz_behavior-driven_2009}, we used \lotse{} to implement suggestions based on interaction pattern detection. Using a VA interface with a scatterplot and a line chart, we defined two strategies: If analysts continuously change the month of data shown in the scatter plot, we suggest using the line chart. An excerpt of this action is shown in \autoref{fig:patterns}. If analysts inspect more than five data points in a row by opening their tooltips, we suggest zooming in and prepare a summary of the data. 
To enable this scenario, we store a sequence of user interactions into \lotse{}'s analysis state. Whenever analysts change the slider or hover over a data point, we add an event to the interaction history. Both strategies are permanently active, and their actions define sequence rules on the tail of the interaction log to determine whether they should produce suggestions in a given situation.

\subsubsection{Goal-Driven Suggestion Adaptation}
\begin{figure}[t]
    \centering
    \includegraphics[width=\linewidth]{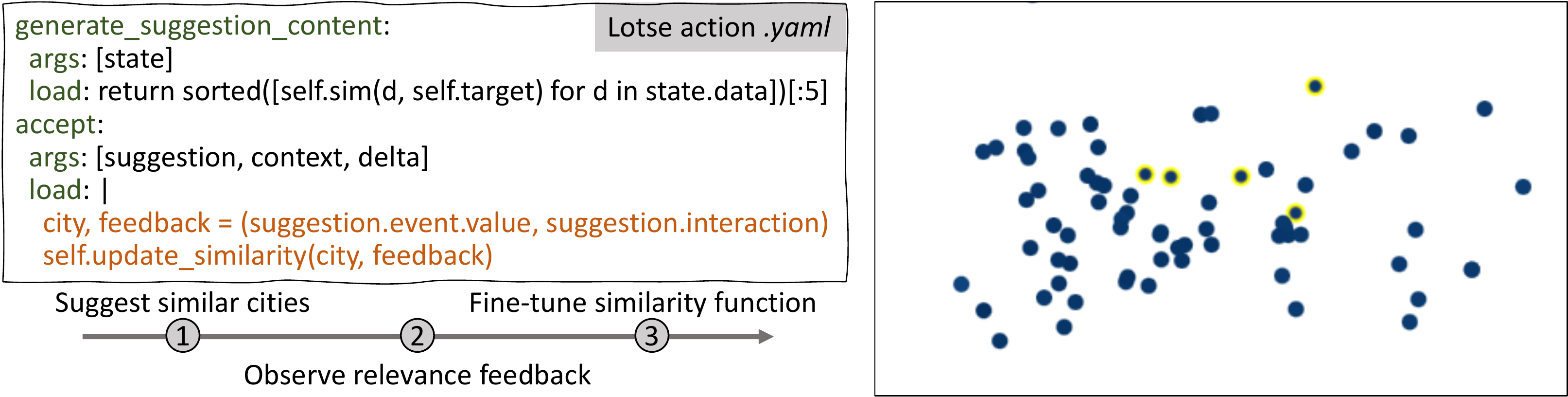}
    \caption{Strategy example in \lotse{} that uses relevance feedback to update a similarity function to guide exploratory data analysis. }
    \label{fig:similarity}
    \vspace{-1em}
\end{figure}

Starting from the same visual analytics interface containing a scatter plot and a line chart, we imagine a task in which analysts have to investigate weather data and identify their favorite city for their next vacation. In the process, analysts select their favorite cities and hide uninteresting cities. Our guidance goal is to support analysts by learning the perceived feature importance for each weather dimension and use it to personalize a weighted similarity function between cities. To that end, we devise two strategies: the first suggests investigating similar cities and  highlights them in the scatter plot. The second strategy suggests switching to a different month if more cities with similar properties to the current favorites can be found there. Accepting or rejecting a suggestion increases (or decreases) the importance scores of all dimensions depending on the similarity between the suggestion and the currently selected favorite cities. All importance scores are stored in the analysis state vector to easily share them between both strategies. To facilitate analysts finding their perfect vacation target (both in terms of location and time), we prioritize month-switching suggestions over similarity highlighting and block the latter if similarity highlighting suggestions are currently available using a meta-strategy. As analysts mark more and more cities as favorites or remove them, they receive new suggestions based on the fine-tuned similarity function. The resulting \lotse{} suggestion workflow is summarized in \autoref{fig:similarity}.

\subsection{VA Expert Case Studies}
In this section, we present two case studies from projects that use \lotse{} to provide guidance. Our case studies aimed to evaluate the usefulness of \lotse{} to VA experts with no experience in guidance design and investigate how they used the library. Below, we report results from both studies individually, in increasing order of guidance complexity and level. 

\noindent\textbf{Methodology --} Both case studies were carried out over one week and started with an introduction session ($\sim1$ hour) in which we presented the library using the demo implementation available on GitHub (code snippets from \autoref{sec: lotse}). We described the idea of creating guidance specifications in yaml files and went through two example strategies to showcase the interplay of strategies, actions, and applicability conditions. In the following week, participants derived and defined guidance strategies and implemented the required frontend changes. During the week, we were in contact with participants to answer clarifying questions and check on their progress. In one case, we also actively supported participants in identifying a syntax error in one of their action files. We concluded the case studies with semi-structured interviews to capture feedback on the experience of using \lotse{}. In particular, we asked participants to summarize their strategies and provide rationales for their design decisions. Finally, we asked participants to evaluate \lotse{} in terms of eight criteria (expressiveness, creativity support, flexibility, efficiency, usability, learnability, and integration into existing workflows) collected by Ren et al.~\cite{ren_reflecting_2018}. 

\noindent\textbf{Participants --} All participants are VA researchers currently designing or implementing VA systems. All participants (two for each case study) have rich experience in system design and development but had not implemented guidance in their systems before. One participant is a final-year master's student, and three participants are graduate students.

\subsubsection{Neural Network Debugging}
\begin{figure}[t]
    \centering
    \includegraphics[width=0.9\linewidth]{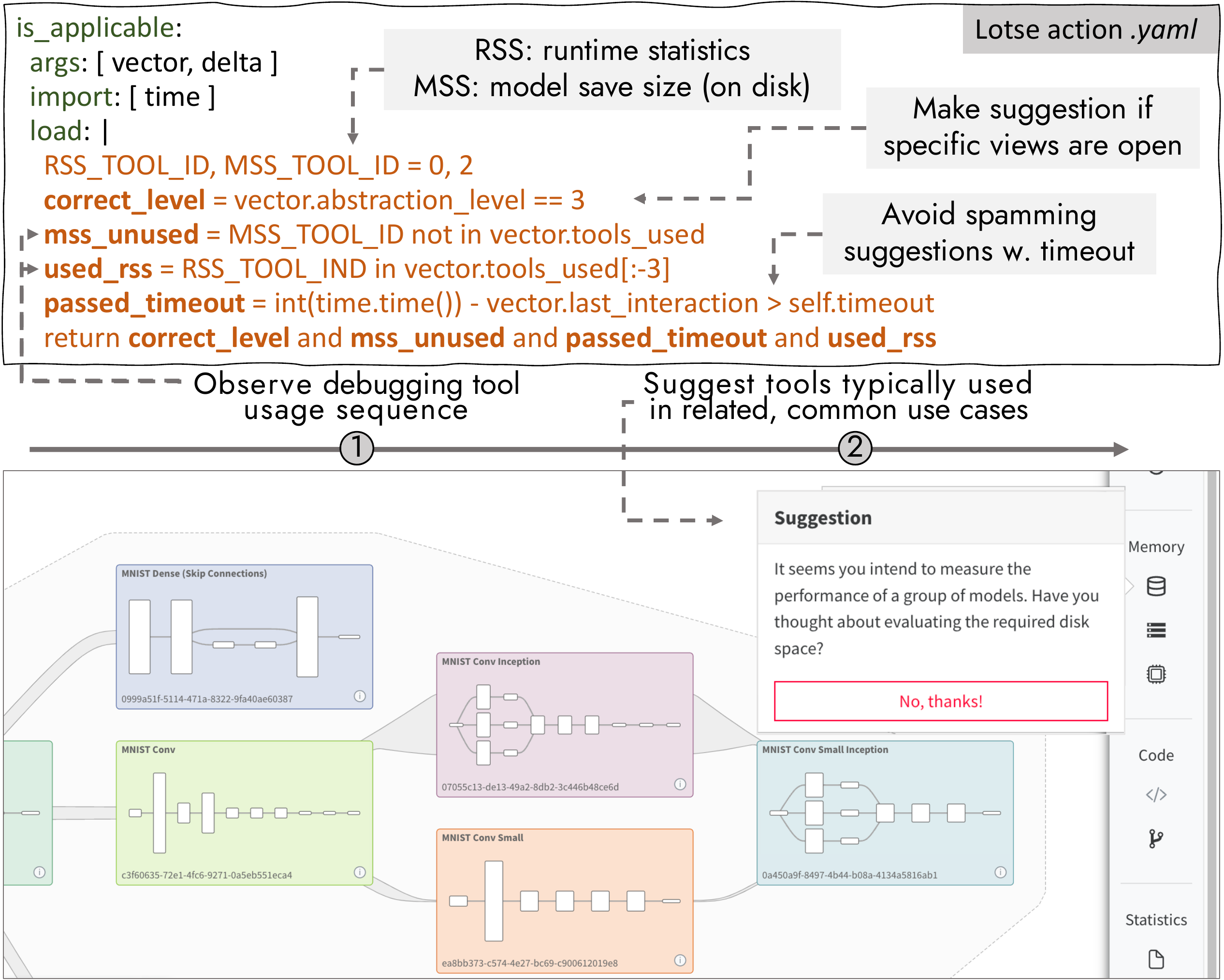}
    \caption{Case study summary for \emph{iNNspector}. The guidance is centered around suggesting alternative debugging tools to explore.}
    \label{fig:innspector}
    \vspace{-1em}
\end{figure}
The first case study was performed in the context of a VA system for structured neural network debugging called \emph{iNNspector}. The system is currently being developed by our participants, and targeted at model developers aiming to understand their models, compare modeling alternatives, and identify potential issues. It supports tracing model behavior to the responsible components (e.g, layers, units, functions) and contains a myriad of \emph{tools} that can be applied to model components, e.g., to show performance curves or inspect activations. 

\paragraph{Goals and Implementation} 
The researchers conceptualized four different approaches to implementing guidance but ultimately decided to focus on guiding their users through the vast amount of debugging tools in the system. In particular, they saw the biggest value in recommending which tools users should investigate next. To that end, they collect the currently selected application abstraction layer, the sequence of previously used tools, as well as the time since the last user interaction in the analysis state vector. An excerpt of the action suggesting to use the \emph{MSS} (model save size) tool is shown in \autoref{fig:innspector}. It first checks that the \emph{MSS} tool has not been used already, but that users recently employed the \emph{RS} (runtime statistics) tool and that the user has not interacted for a while. A large collection of such actions can be used to introduce potentially relevant debugging tools to users quickly. As the participants had common tool usage sequences in mind to solve different tasks, they did not implement adaptation or the contextualization to user feedback.

\paragraph{Usage Experience}
The participants reported that implementing guidance using \lotse{} was ``\emph{not difficult}'' but ``\emph{efficient}'' and ``\emph{intuitive}'', especially as the ``\emph{definition of strategies is straight-forward}'' while providing significant freedom in the definition of callbacks. While they praised the freedom of \lotse's callbacks, they also stated that ``\emph{defining strategies as yaml files brings a lot of flexibility}'' and was a ``\emph{good way of declaring what you want}'', similar to writing docker-compose files or Kubernetes manifests. However, they also stated that \lotse{} did not feel as event-driven as they might have liked as there was no way to subscribe to partial state vector updates. Further, they requested to inject new, dynamic strategies from the visualization frontend as required. We plan to address both comments in future releases of \lotse{}. 

\subsubsection{Music Sheet Corpus Analysis}
\begin{figure}[b!]
    \centering
    \vspace{-1em}
    \includegraphics[width=\linewidth]{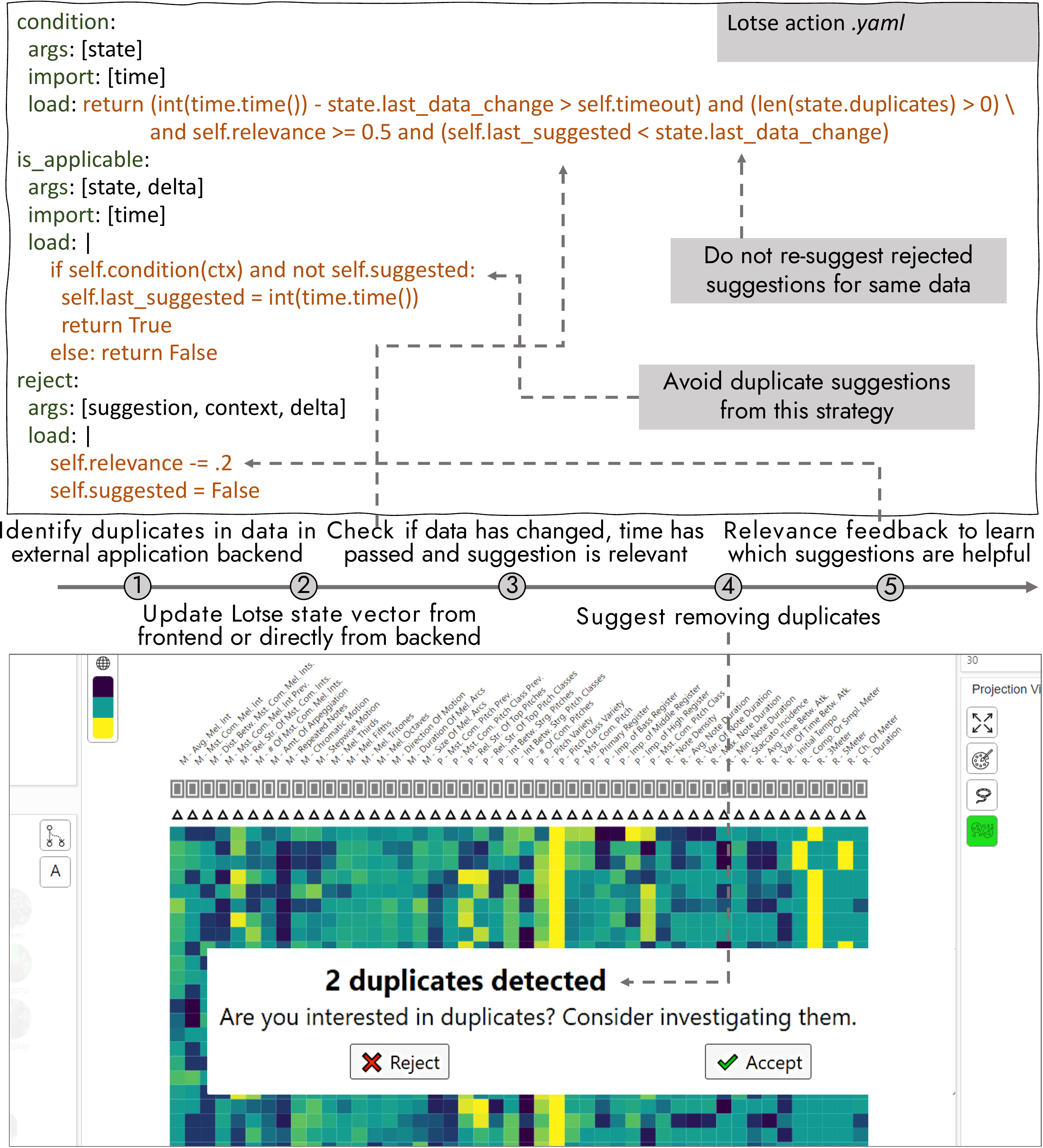}
    \caption{Case study summary for \emph{CorpusVis}. The six provided guidance strategies focus on prescribing suggestions for data exploration and visualization interactions.}
    \label{fig:corpusvis}
\end{figure}

We performed the second case study with researchers who had just finished the implementation of \emph{CorpusVis}, a VA system for music sheet corpus analysis~\cite{miller_corpusvis_2022}. The results are summarized in \autoref{fig:corpusvis}.

\paragraph{Goals and Implementation} 
In CorpusVis, users can explore corpora of sheet music using a set of coordinated views. The researcher's overarching guidance goal was to support this data exploration with a mixture of suggestions on the \emph{data exploration} and \emph{analysis interaction} levels. In total, they implemented six strategies targeting duplicate removal, data aggregation, or the addition of more potentially relevant data to the analysis, among others.  As \emph{CorpusVis} already uses intro.js in the onboarding process,  suggestions are also shown via this framework. As a result, all suggestions are prescribing, and at most one suggestion is shown at a time. The example action from \autoref{fig:corpusvis} suggests the removal of duplicate data from analysis. Potential duplicates are identified in the CorpusVis system and sent to \lotse{} via its state update API. The strategy then follows a pattern the researchers frequently used: It stores a relevance score which is increased or decreased upon acceptance or rejection of suggestions from that strategy, respectively. If a strategy's relevance falls below a threshold, suggestions from that strategy are temporarily disabled. Further, the researchers conceptualized various approaches for contextualizing the various strategies as they ``\emph{begin to compete with each other.}'' They finally implemented conditional dependencies between strategies, where some strategies only activate if suggestions from others were recently accepted or rejected.

\paragraph{Usage Experience}
As ``\emph{it does not much take time to integrate in your system}'' these participants used \lotse{} extensively and were able to achieve results quickly. They found that ``\emph{by implementing one strategy you immediately have more ideas of what you would also like to do}'', validating our goal to make reasoning about guidance implementation more approachable. They found that designing guidance with \lotse{} ``\emph{helped to structure the process}'' and provided a starting point for guidance implementation that they were missing before.  While they also praised that \lotse{} did not restrict them in the guidance they wanted to implement, they sometimes found it difficult to debug the contextualization of their suggestions. Nonetheless, they reported that they did not need much time to get used to implementing in \lotse{} and stated that ``\emph{it is worth it as it can be reused in [their] future projects}''. They further appreciated the definition of strategies and actions in yaml files as they provided ``\emph{a clear structure and overview of your strategies}''. Overall, they found the integration of guidance using \lotse{} ``\emph{very beneficial for [their] application}'' and expressed the wish for ``\emph{a gallery of examples to learn from and to adapt}'' that we plan to create together with the community. 

\section{Discussion}
In this paper, we have proposed an implementable model of adaptive guidance in VA centered around guidance templates and strategies. This model bridges the gap between complex, often impractical theoretical models and what is possible to implement today.  We demonstrate this practical implementability with a library that generates running guidance engines from strategy- and action-specifications in yaml files.

\subsection{Lessons Learned}
\paragraph{Guidance Levels} 
While we originally designed \lotse{} to cover exploratory and task-specific guidance, developer feedback showed that \lotse{} can also be used to contextualize onboarding-like guidance. Rather than relying on static onboarding tours, \lotse{} enabled them to provide contextualized onboarding cues and adapt them to user feedback. From our observations, the guidance level determines whether the focus of change in strategy-based guidance is on contextualization (i.e., providing suggestions in different situations) or adaptation (i.e., varying the suggestion content).  

\paragraph{Guidance Implementation Beyond Strategies} 
As a direct consequence of this observation, we do not claim that all guidance in the future should be generated from  strategies following this model. We do, however, believe that strategies are an appropriate vehicle for novice  designers, excel in multi-objective optimization problems where different objectives can be expressed as individual strategies, and when users frequently switch between exploration and specific tasks. \lotse{}, as a particular instantiation of strategy-based guidance, is similarly targeted towards novices implementing simpler guidance and might be insufficient for very complex guidance generation processes. 

\paragraph{Contextualization of Guidance} 
By making analysis state vectors first-class citizens, \lotse{} forces developers to think about contextualization of guidance. While previous work has shown that helpful guidance can also be provided without explicit contextualization, adapting guidance to the analysis context holds the promise of a more effective human-machine collaboration. Thinking about guidance contextualization in terms of composable rules rather than fixed ``if the user clicks here, then...''-conditions enabled the participants in our case study to come up with intricate  strategies. 

\paragraph{Explainable and Justified Guidance} 
Explainability of suggestions and their creation is a common challenge in guidance, with users missing explanations (how?) and justifications (why?) of guidance they received. We integrate description fields into each generated suggestion. Hence, developers can either provide a description or actively decide not to do so. We hope that by raising awareness for the importance of guidance justifications beyond traditional XAI methods, new approaches for justifiable guidance will emerge. 

\subsection{Limitations}
Our approach of strategy-based guidance assumes that all information that is necessary to decide which guidance to provide next can be stored in an analysis state vector. In theory, it is possible to store data like facial expressions~\cite{ceneda_show_2021} or eye-tracking results into the analysis state. However, \lotse{} is arguably built for simpler data like relevance feedback, mouse positions, or interaction timings and sequences. We believe that this data already provides a rich foundation for guidance provisioning and is successfully used in many guidance applications today.

While we set out to make guidance implementation easier and more approachable for developers without previous guidance experience, \lotse{} still has a learning curve. As it starts a python web server to produce suggestions, its barrier to entry is higher than that of fully browser-based approaches. Also, we opted to allow imperative callbacks in \lotse{} rather than fully restricting it to a declarative grammar. As our case studies only feature four participants, further observations are necessary to judge the impact of our decisions when designing \lotse{}.

When discussing prototypes of \lotse{} with colleagues, we sometimes met the expectation that the library could provide guidance for their bespoke VA systems out-of-the-box. However, \lotse{} does not contain some generic pre-trained ML models that could be generically employed. Instead, developers must design strategies and adapt existing visualizations to synchronize analysis state vectors. The current version of the library also does not offer a gallery of ready-to-use examples with broad applicability. Creating such a gallery would facilitate setting up guidance even further and is thus planned for the near future.

\subsection{Opportunities \& Next Steps}
Our efforts in implementing an approachable guidance library have revealed several opportunities for future work.

\paragraph{A Guidance-Grammar for VA} 
Visualization grammars have seen a large increase in popularity lately, with newer grammars tackling more and more complex issues such as animations~\cite{kim_gemini_2021} or responsive visualization~\cite{kim_cicero_2022}. Our implementation provides a first step towards a general, declarative \emph{guidance grammar}. While it already supports automatic conversion of created strategy templates into code and offers some declarative features, it is also reliant on imperative callbacks. Identifying if and how the vast space of available and potential guidance systems could be described using a declarative approach--potentially in close integration with a visualization grammar like Vega-Lite--opens an exciting direction for future research.

\paragraph{Defining Guidelines} 
As we move towards more readily available guidance implementations, providing guidance will become more approachable to  novices. As a result, the definition of guidelines that capture and mitigate common issues such as bias, agency, or simply alienating the user is necessary. The aforementioned creation of a gallery with prominent  strategy examples constitutes a first step towards deriving reliable, practical guidelines. 

\paragraph{Guidance Generation Inputs}
As mentioned above, a current limitation of \lotse{} stems from its focus on analysis state data that can be captured by observing interactions on the frontend. However, other inputs such as webcam video, sound, or eye-tracking have also been used to generate guidance in the past. Comparative evaluations could identify in which contexts such complex features are necessary and when simpler input features are equally effective.

\paragraph{Capturing User Feedback} 
At the core of our method is the choice and evaluation of multiple strategies. To employ these strategies most effectively, it is necessary to capture and model user feedback. This feedback can then be employed to adapt the guidance, utilizing strategies considered relevant. For specific model-optimization tasks, a wide variety of relevance-feedback mechanisms have been proposed in the past. Future work is needed to determine which candidates are most promising to capture guidance feedback and initiate adaptation. 

\paragraph{Partial Automation}
Learning contextualized user preferences on targeted guidance strategies through relevance feedback holds the promise of enabling partial automation of suggestions in specific contexts. While there are apparent concerns regarding the locus of control and agency, the participants from the musicology case study also expressed interest in exploring this direction. Splitting the available guidance into strategies allows learning if and when to automate specific guidance and enables the definition of specific learning goals~\cite{sperrle_co-adaptive_2021} for each  strategy that need to be reached before automation can take place. 

\section{Conclusion}
In this paper, we summarized existing theories of guidance in VA and derived a model of strategy-based guidance that is closer to actual implementation practices. To demonstrate the usefulness of our model, we have instantiated it in a guidance library called \lotse{}. \lotse{} enables the specification of guidance and automatically generates running code. It is available at \href{https://github.com/lotse-guidance}{\texttt{https://github.com/lotse-guidance}}.  In the future, we will continuously improve and extend \lotse{} with a gallery of predefined guidance strategies and work towards further facilitating the creation of strategies. 

\acknowledgments{
This work was funded by Vienna Science and Technology Fund (WWTF) under grant ICT19-047, by Deutsche Forschungsgemeinschaft (DFG) under grant  455910360 (SPP-1999), and the ETH AI Center.}

\bibliographystyle{abbrv-doi-hyperref}
\bibliography{vis2022}

\end{document}